# Very Efficient Spin Polarization Analysis (VESPA): New Exchange Scattering-based Setup for Spin-resolved ARPES at APE-NFFA Beamline at Elettra

Authors

**Chiara Bigi**[ab], **Pranab K. Das**[b], **Davide Benedetti**[b], **Federico Salvador**[b], **Damjan Krizmancic**[b], **Rudi Sergo**[c], **Andrea Martin**[b], **Giancarlo Panaccione**[b], **Giorgio Rossi**[ab], **Jun Fujii**[b]* and **Ivana Vobornik**[b]*

[a]Department of Physics, Università degli Studi di Milano, Milano, I-20133, Italy

[b]CNR-IOM Laboratorio TASC, Trieste, I-34149, Italy

[c]Elettra Sincrotrone Trieste, Trieste, I-34149, Italy

Correspondence email: fujii@iom.cnr.it; vobornik@iom.cnr.it

**Abstract** *Complete Photoemission Experiments*, enabling to measure the full quantum set of the photoelectron final state, are in high demand for the study of materials and nanostructures whose properties are determined by strong electron and spin correlations. We report here on the implementation of the new spin polarimeter VESPA (Very Efficient Spin Polarization Analysis) at the APE-NFFA Beamline at Elettra that is based on the exchange coupling between the photoelectron spin and a ferromagnetic surface in a reflectometry setup. The system was designed to be integrated with a dedicated Scienta-Omicron DA30 electron energy analyzer allowing for two simultaneous reflectometry measurements, along perpendicular axes, that, after magnetization switching of the two targets allow to perform the 3D vectorial reconstruction of the spin polarization while operating the DA30 in high resolution mode. VESPA represents the very first installation for spin resolved ARPES (SPARPES) at the Elettra synchrotron in Trieste, and is being heavily exploited by SPARPES users since fall 2015.

## 1. Introduction

Probing the spin-resolved electron states of solids, surfaces and nanostructures gives direct access to phenomena like magnetism, proximity effects, spin-orbit interaction and related spin texture in low dimensional systems (Bianchi *et al.*, 2010; Jozwiak *et al.*, 2011; Bahramy *et al.*, 2012; Riley *et al.*, 2014; Suzuki et al., 2014; Veenstra *et al.*, 2014; Das *et al.*, 2016; Mo *et al.*, 2016). In particular, spin-orbit coupling manifests with the Rashba effect in 2D systems (Hoesch *et al.*, 2004; Yaji *et al.*, 2010; Sakamoto *et al.*, 2013; Takayama *et al.*, 2014), Weyl semimetals (Lv *et al.*, 2015; Yang *et al.*, 2015) and in the topological insulators where the surface states are characterized by a chiral spin texture and the spin is locked to the electron momenta. (Hsieh *et al.*, 2009; Xia *et al.*, 2009; Pan *et al.*, 2011; Zhu *et al.*, 2014) Such "quantum materials" are regarded as potential functional systems for developing low energy fast spintronic sensors or logic devices. Increased interest in the electronic properties of these new materials, witnessed also by the recent Nobel awards in physics, is reflected in increased demand for Spin-polarized Angularly Resolved Photo-Electron Spectroscopy (SPARPES or Spin-ARPES). The full implementation of SPARPES requires the energy and polarization control of the incoming photon beam (photon-in, from synchrotron radiation, laser or UV-lamp sources) and the simultaneous energy, momentum and spin polarization analysis of the ejected photoelectrons (electron-out; see, for example, Hüfner (2003), Damascelli (2004), Osterwalder (2006), Seddon (2014) and references therein). In the non-interacting approximation, the kinetic energy ($E_{kin}$) and the emission angle ($\vartheta$) of the photoelectrons are straightforwardly related to the binding energy ($E_{bin}$) and the vector momentum $k_{//}$ (parallel to the sample surface)[1] of the initial bound electron state respectively. Therefore, analysis of the photoemission current provides access to the initial state, as sketched in figure 1. Moreover, the spin orientation is not affected by the dipolar interaction during the photoemission process. When the spin polarization (SP) of selected bands is a key information, the full characterization of the electron states can therefore be achieved if all the quantities ($E_{kin}$, $\vartheta$, SP) are simultaneously measured, and furthermore the photoionization cross sections and symmetry selection rules can be exploited by properly choosing the photon energy and the polarization state of the radiation to control the experimental geometry. While the high resolution analysis of photoelectron energy and momentum, has become widely available thanks to the advances in electrostatic condenser type or time of flight type analyzers, the full vectorial determination of the spin polarization of the photoelectron is still limited by the statistics of high resolution photoemission experiments. The main limiting factor in Spin Polarization measurements based on Mott scattering experiments is the low cross section for spin-orbit scattering and the resulting poor figure of merit (FOM) of such detectors (Gay & Dunning, 1992). Indeed, the typical Mott detector efficiency of ~$10^{-4}$ has hindered the spin analysis of high resolution ARPES. Other approaches have therefore been developed, which exploit the exchange interaction between low energy photoemitted electrons and magnetically ordered target surfaces in LEED-like reflectometry geometries, leading to the increase in efficiency by a factor $10^2$ and rendering therefore this type of detection suitable for high resolution ARPES (Kutnyakov *et al.*, 2013; Jozwiak *et al.*, 2010; *Bertacco et al.*, 2002). Here we report

---

[1] It is possible to retrieve the component perpendicular to the surface $k_\perp$, which is not conserved in the photoemission process, if suitable assumption on the photoelectron final state are provided (Hufner (2003), Damascelli (2004))

on the design, the implementation and the operational parameters of an optimized 3D vectorial spin polarimeter that operates in the *Very Low Energy Electron Diffraction* (VLEED) regime. The system has recently become available as users' station at the APE-NFFA Beamline of IOM-CNR at the Elettra synchrotron radiation facility in Trieste (Italy).

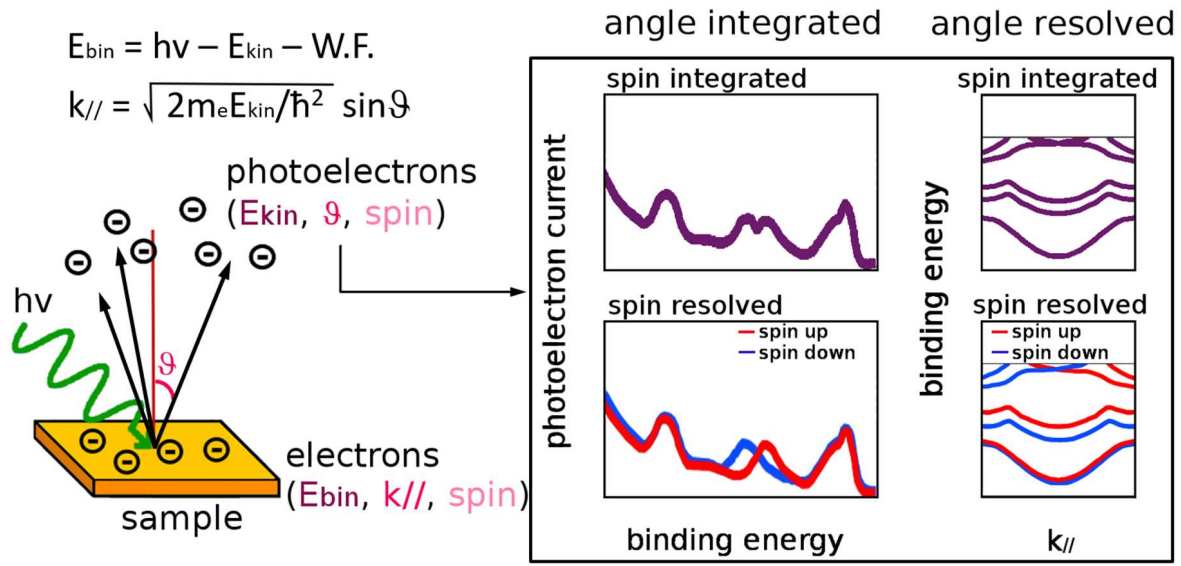

**Figure 1** Left: Sketch of the photoemission process. The quantum numbers of both bound and photoemitted electrons are pointed out together with the relations linking them. Right: Depending on the capability to filter one or more photoelectron quantum numbers, different types of analysis are possible.

## 2. Detector choice and experimental geometry

The VESPA apparatus is at the core of the novel end-station of the Low Energy branch of the APE-NFFA beamline, operating with a dedicated quasi periodic APPLE-II undulator in the near and far ultraviolet regime (photon energy range 10-100 eV) with full polarization control and high spectral purity (Panaccione *et al.*, 2009). The photoemitted electrons are collected by a DA30 hemispherical electron energy analyzer[2], operating in the deflection mode within up to 30° solid angle acceptance for a given experimental geometry. The spectrometer can be operated routinely at 10 meV energy resolution (overall including synchrotron radiation bandwidth) and it allows to acquire energy distribution curves and momentum distribution curves with high efficiency from an illuminated sample area of 50x150 µm$^2$, ARPES band mapping enables to identify the relevant bands in the Brillouin zone and apply the spin analysis at fixed experiment geometry, i.e. without changing the sample polar angle therefore at (almost) constant matrix elements and facilitating measurements on small samples (i.e. sample grains < 500x500 μm$^2$).

---

[2] Detailed information on the Scienta-Omicron DA30 analyzer is available at the company website: http://www.scientaomicron.com/en/products/da30-arpes-system/1280

The scheme of our setup is shown in Figure 2. The VESPA spin polarimeter is made of two orthogonal VLEED reflectometers receiving two energy and momentum selected electron beams from the exit plane of the DA30 analyzer. The design was inspired by the ESPRESSO (Efficient Spin REsolved SpectroScopy Experiment) setup installed at the beamline BL-9B of the Hiroshima Synchrotron Radiation Centre (Okuda *et al.*, 2011; Okuda *et al.*, 2015). In our case, the 80mm diameter exit of the DA30 analyzer (Figure 2(b)) is fitted with a 40 mm MCP detector for spin-integrated, high resolution ARPES and with two apertures, aligned along the kinetic energy dispersion axis, embracing the MCP. Through such apertures, the photoelectrons are channeled, transferred and focused by an eight-element electrostatic lens onto the exchange scattering targets in the twin VLEED spin polarimeters. Two identical VLEED scattering chambers, each receiving energy and momentum filtered photoelectrons from the respective aperture, are oriented at 90° with respect to each other in order to measure the three SP components of the photoemitted electron beam, allowing for vectorial reconstruction of the SP. The 3D rendering of the all μ–metal UHV apparatus is shown in Figure 2(c). The targets are magnetized in the UHV scattering chambers by two pair of solenoids (applied field 100 Oe) that impose two orthogonal magnetization axes in the target surface. We measure the (00) diffracted beam from the magnetic target and determine the spin asymmetry between two subsequent measurements obtained with two antiparallel target magnetization directions. The arranged geometry provides that the VLEED-W polarimeter measures the spin component aligned parallel to the analyzer slit (the APE DA30 momentum dispersion plane is perpendicular to the Elettra orbit), projecting on the vertical x-axis within the sample surface plane in the experimental setup of Figure 2(a), i.e. the x in-plane spin polarization (in the condition of normal emission); the other component is perpendicular both to the analyzer slit and to the sample surface (z out-of-plane component). The second polarimeter, VLEED-B, gives the spin polarization components along the direction parallel to the sample surface and perpendicular to the slit (in-plane y component) as well as the same out-of-plane (z) component that is redundant with the identical measurement by VLEED-W. The possibility to measure the same z polarization component with both W and B polarimeters (i.e. two different targets) allows to cross-normalize the data and retrieve the true vectorial information on SP.

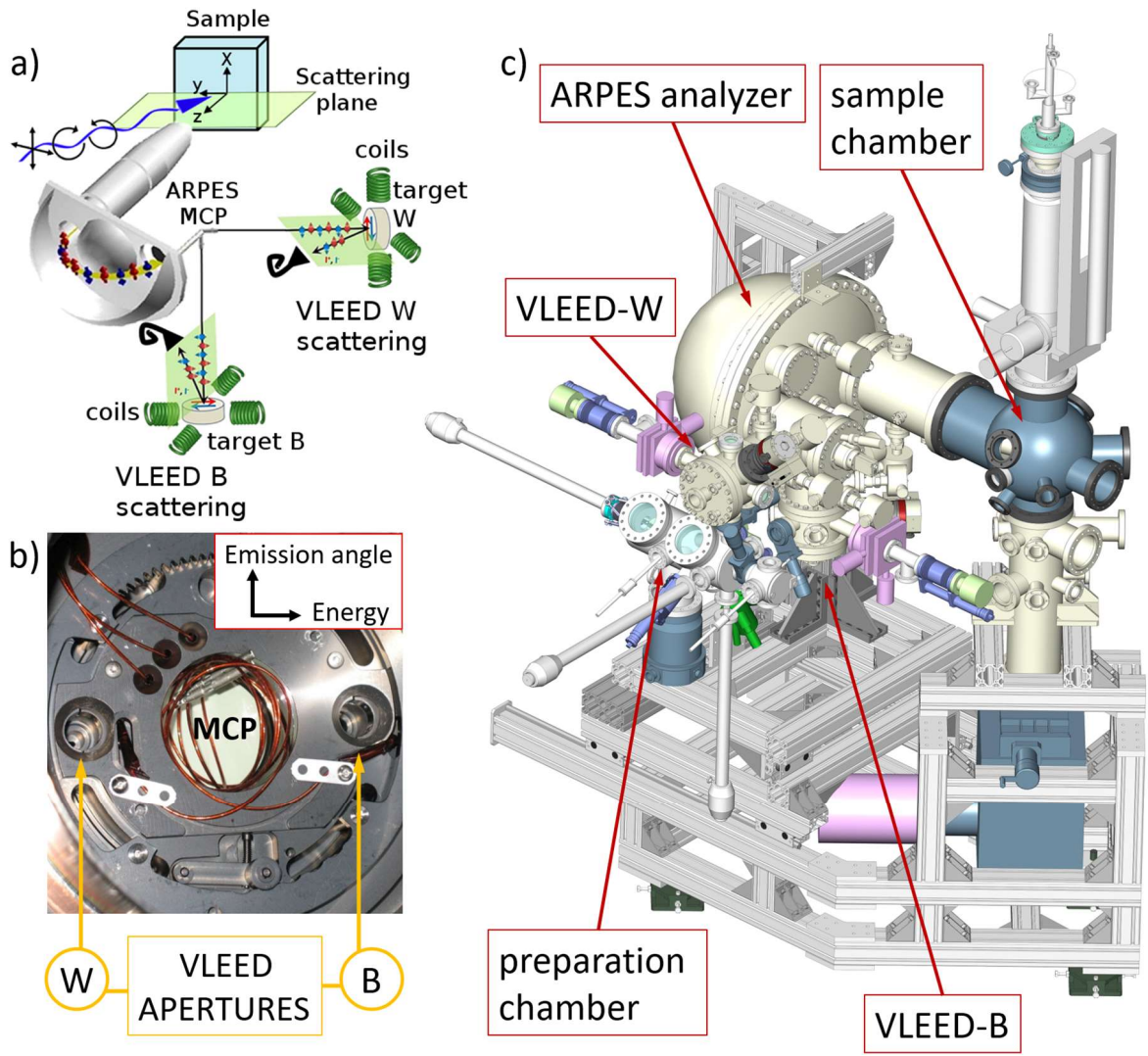

**Figure 2** a) Experimental geometry scheme. Polarized monochromatic synchrotron radiation is travelling at 45 degrees with respect to the analyzer lens axis, in the horizontal plane of the storage ring, while the analyzer slit is perpendicular to it. (b) The detector setup placed on the analyzer exit plane with central MCP and two apertures along the energy dispersion axis through which the selected electrons are sent into the spin polarimeters (aperture W/B lead electrons to VLEED-W/VLEED-B respectively); the size of the aperture (four possibilities) can be selected via a carousel (not visible in the figure) and determines the energy and momentum resolution as described in Okuda et al., 2011. (c) 3D rendering of the APE-LE end-station, the labelling shows all the sections of the VESPA analyzer as well as the sample chamber and the preparation chamber, which is directly connected to the VLEED scattering chambers.

The size of the aperture of the spin channel determines the angular and the energy resolution of the spin polarimeter: four sets of apertures are selectable in situ (apertures areas are 0.5x1mm$^2$, 1x2mm$^2$, 2x3mm$^2$, 4x4mm$^2$; the first value is the size along the angular dispersion axis while the latter is the size along the energy dispersion axis). The overall detector resolution is obtained by the convolution between the chosen aperture size, the electron pass energy through the hemispherical condenser, the lens mode and the entrance slit aperture of the hemispherical analyzer (see table II in Okuda *et al.*, 2011).

The set of transport and focusing electrostatic lenses leads the monochromatic electron beam to the target and allows to accelerate/decelerate to the optimal exchange scattering energy, that for our target is 6 eV. This has implications on the design of the lenses and polarimeter as stray magnetic fields or parasitic electrostatic fields have disruptive effects on low energy spin polarized electron beams. Minimizing the external magnetic field is thus mandatory for spin measurements. For this reason, the UHV vessels and the entire components inside the VESPA scattering chambers are made with non ferromagnetic materials (e.g. aluminum and molybdenum) and are enclosed by magnetic shields made of thick µ–metal foils. The target for electron scattering is fixed on a Mo plate that can be easily manipulated during the in-situ target preparation (in the dedicated separate UHV chamber in between the two scattering chambers; white color in Figure 2c) and further inserted in the target manipulator of the polarimeter chambers via a UHV transfer. The target manipulator has four degrees of freedom: x, y, z and ϑ. Two orthogonal pairs of kapton insulated copper coils surround the target and generate the pulse of magnetic field for target magnetization (Figure 3). As sketched in figure 3, each pair has its axis along one of the two in plane easy magnetization directions of the iron target (i.e [100] and [010] axis). A short pulse of current (5A) generates ~100 Oe magnetic field that magnetizes the target. Measurements are then performed with the target in remanent magnetization, and zero applied field.

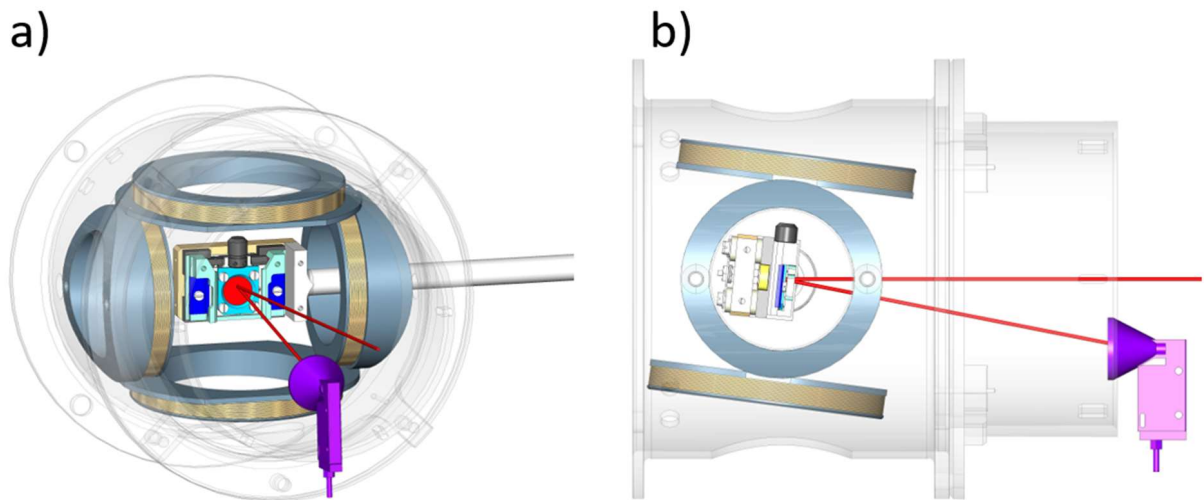

**Figure 3** sketch of the experimental geometry in the VLEED scattering chamber. The ferromagnetic target is shown in red color as well as the path of the photoemitted electrons (both incoming and scattered) the two pair of coils (gray colored) for target magnetization are also shown togethers with the µ-metal shield surrounding the scattering zone (a) front view (b) side view.

The scattering target consists of a ~1200 Å thick epitaxial iron film grown in situ on MgO(001). The target surface is stabilized by a controlled oxidation procedure leading to the formation of a well ordered (1x1) FeO passivated surface, while excess oxygen is desorbed. Such films are characterized by high remanent magnetization, low coercivity, and large exchange asymmetry (Bertacco & Ciccacci, 1999; Bertacco *et al.*, 1999) for absorption/reflection of electrons at kinetic energies in 5-15 eV range. The exchange coupling at these electron energies is much stronger than the spin-orbit coupling at 20-200 keV yielding to a significant increase in spin discrimination efficiency (larger by two orders of magnitude) that we will discuss below.

The oxygen stabilized iron target has a long lifetime in UHV conditions: the detector performance degrades by ~10% in two months of operation at $5·10^{-11}$ mbar base pressure. Furthermore, the pristine state of the target surface can be easily recovered implementing a quick and reliable annealing protocol (Bertacco *et al.*, 1998). Similar targets have been also employed in other exchange scattering spin polarimeters, including the ESPRESSO setup. (Okuda *et al.*, 2015; Winkelmann *et al.*, 2008; Hillebrecht *et al.*, 2002; Okuda *et al.*, 2008; Bertacco *et al.*, 1999).

As mentioned earlier, VESPA includes a dedicated target preparation chamber with UHV transfer to both VLEED-B and VLEED-W. The target preparation chamber is fit with an e-beam deposition source of ultrapure iron onto in situ annealed MgO(001) substrates, and a LEED diffractometer for in situ diagnostics of the surface quality. The LEED pattern in Figure 4 confirms that the bcc iron targets grow in single domain and without surface reconstruction. This quality is crucial to have reliable spin asymmetry measurement, avoiding spurious scattering from misoriented magnetic domains. The Fe(001)-p(1x1)O surfaces have identical in-plane easy magnetization axes at 90-degrees and the targets are mounted with those axes aligned with the two sets of solenoids used to magnetize in either of the two easy directions. For this reason, the two targets (W and B) mounted orthogonal to each other as in Figure 2a allow to measure the 3D vectorial components of the spin polarization.

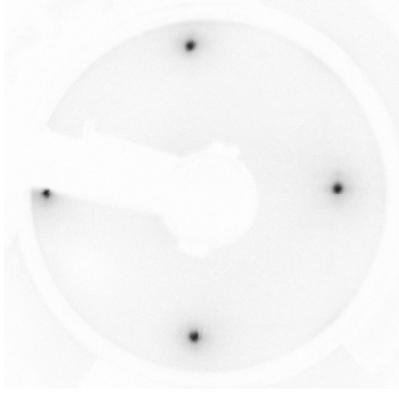

**Figure 4** LEED pattern of the oxidized Fe target measured with electron kinetic energy $E_k$ = 90 eV indicating the single domain ordered film surface.

## 3. Performance

We tested the performance of the VESPA polarimeters by measuring the Rashba-split surface states of Au(111), as a spin polarization standard (LaShell *et al.*, 1996) and the spin-polarized surface state of the $Bi_2Se_3$ topological insulator (Hsieh *et al.*, 2009).

### 3.1. Au(111) Schockley state

The Au(111) surface presents a parabolic surface state in the L-gap of the bulk band structure, whose electrons can be regarded as forming a free electron gas propagating in two-dimensions. The spin-orbit coupling lifts the spin degeneracy at the surface generating two electron surface states that are fully spin-polarized as predicted by the theory of Rashba (Bychkov & Rashba, 1984; Henk *et al.*, 2003) and whose spin texture was probed by SPARPES with Mott scattering, as in the COPHEE experiment (Hoesch *et al.*, 2004). The spin is lying in the (111) surface plane and aligned perpendicularly to the electron momentum. The energy/momentum separation between the two Rashba-split bands reaches ~110 meV and ~0.5° (i.e. ~0.024 $Å^{-1}$) at the Fermi energy, making the Au(111) surface state a good candidate for probing the effective overall resolution of our spin polarimeter.

The Au(111) surface of a [111] oriented high-purity single crystal was prepared in situ, under UHV conditions, by standard cycles of Ar+ ion sputtering and annealing, resulting in well ordered surface displaying sharp and well resolved Rashba-split surface states (Figure 5(a)). The ARPES data were acquired with 30 eV photon excitation energy and are shown in Fig. 5(a). We selected the position in the Brillouin zone ($k_x$, $k_y$) = (0.096 $Å^{-1}$, 0 $Å^{-1}$) for performing the acquisition of SPARPES. The choice of the aperture (the smallest, i.e. #1) for spin channels combined with the pass energy and the angular lens mode of the analyzer (for this specific case 20 eV and 14° respectively) determines the angular resolution of ~0.38° (the angular window confined within the two red lines in Fig. 5(a)), i.e. ~0.018 $Å^{-1}$ at the chosen photon energy (30 eV), and an overall energy resolution of ~70 meV. Figure 5(b) shows the resulting spin-resolved ARPES spectra (spin resolved electron distribution curves SEDCs) measured with the VLEED-B polarimeter in the in-plane magnetization configuration (spin perpendicular to the analyzer slit, i.e. perpendicular to the angular dispersion direction). The polarization (P) in Fig. 5(c) was calculated using an effective Sherman function $S = 0.5$ (as derived by comparison with the data reported in Hoesch *et al.,* 2004) and using formula:

$$P = \frac{1}{S}\frac{EDC_{m\uparrow} - EDC_{m\downarrow}}{EDC_{m\uparrow} + EDC_{m\downarrow}},$$

where $EDC_{m\uparrow\downarrow}$ are the measured energy distribution curves for the two opposite directions of the target magnetization *m*. The SEDCs in Figure 5(b) were extracted from the measured EDCs as:

$$SEDC_{\uparrow\downarrow} = \frac{(1 \pm P)(EDC_{m\uparrow} + EDC_{m\downarrow})}{2}.$$

The FOM that characterizes the total performance of the polarimeter is defined as $S^2 \cdot I/I_0$, where *I* is the current of the electrons scattered from the target (measured by the VLEED channeltron) and $I_0$ is the current of electrons entering the VLEED aperture. To estimate $I_0$, we counted the number of electrons in the area of the MCP comparable to the VLEED aperture. We obtained $I/I_0 = 0.13$, which means FOM = $3.3 \cdot 10^{-2}$ using the Sherman function of our targets ($S = 0.5$). This is approximately a factor $10^2$ higher than for the Mott polarimeters (Hoesch *et al.*, 2002) and in accordance with the FOM of the previously built VLEED polarimeters (Hillebrecht *et al.,* 2002; Okuda *et al.*, 2011). The robustness reached with VESPA in terms of target production and lifetime, as well as of reliable sequence of target magnetization and data acquisition, makes the overall setup competitive and fully adapted for advanced studies of correlated electron systems. Moreover, as far as the VESPA apparatus exploits a continuous and stable photon source (i.e. the Elettra synchrotron), the accuracy on the spin polarization value is not affected by the fluctuations of the source (they are essentially negligible in case of synchrotron radiation sources indeed), but relies only on the inverse of the FOM. Therefore, the $10^2$ higher FOM that VESPA displays compared to the Mott polarimeters confirms that the choice for a detector based on exchange scattering is the best for our case (see ref. Pincelli *et al.*, to be published for accurate discussion).

Taking advantage of the high FOM of VESPA we measured the spin-resolved data all along the parabolic dispersion of the surface states displayed in Fig. 5(a). The data were measured at relaxed conditions of the angular and energy resolution (the aperture #2 i.e. ~0.8° angular acceptance, corresponding to ~0.38 Å$^{-1}$ at 30 eV photon energy) and the overall energy resolution is ~100 meV (analyzer pass energy 20 eV). This allowed us to obtain the whole set of spin polarized dispersion shown in Figure 5(d) in ~1 hour.

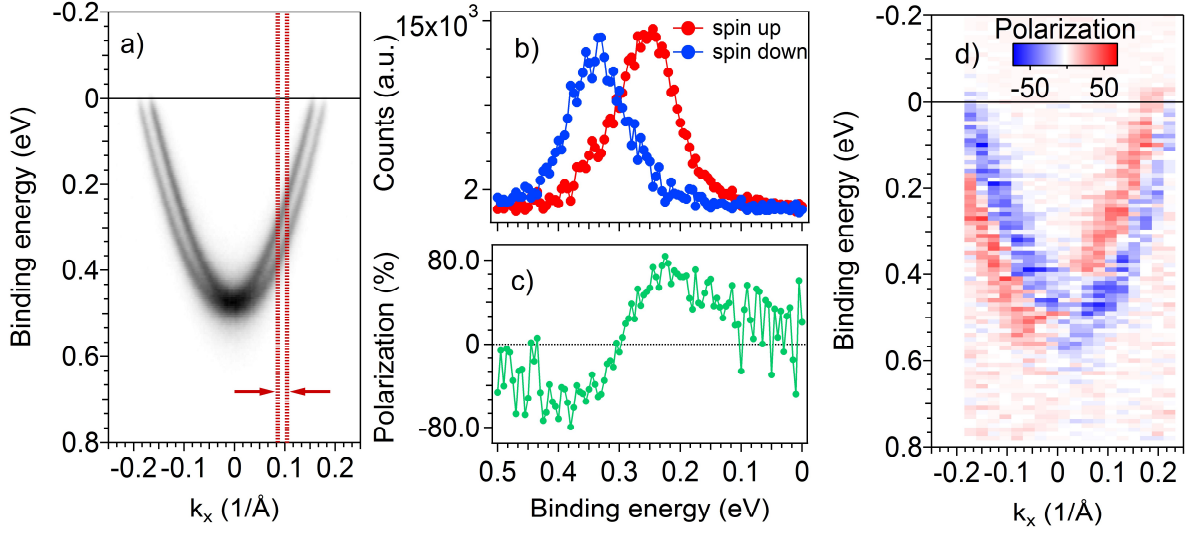

**Figure 5** (a) ARPES data of the Rashba-split Shockley surface state of Au(111); (b) spin-resolved EDCs measured at $(k_x, k_y) = (0.096\ \text{Å}^{-1}, 0\ \text{Å}^{-1})$ with angular resolution of $0.018\ \text{Å}^{-1}$ (within red lines in (a)); (c) the corresponding spin polarization; (d) the spin resolved ARPES map showing the spin polarization (blue-red scale) of the Rashba states. All the measurements were performed at hν=30 eV photon energy.

### 3.2. Bi$_2$Se$_3$ single crystal surface (cleaved in UHV)

The performance of the polarimeter was further tested by measuring the topologically protected surface states on Bi$_2$Se$_3$ topological insulator, where due to the spin-momentum locking the two branches of the Dirac cone are spin polarized with antiparallel spins (Pan *et al.*, 2011). The ARPES signal from the topological surface states is weaker (by a factor ~10) with respect to Au(111) surface states. All the measurements were performed at hν = 50eV photon energy; the ARPES data of Figure 6(a) show the linear dispersion features of the single Dirac cone situated at the center of the surface Brillouin zone. The spin-resolved ARPES data are reported in the colored inset of Figure 6(b). The data clearly reflect the properties of the topologically protected surfaces whose spin helical texture is linked to the crystal momentum, resulting in antiparallel polarizations of the opposite branches of the Dirac cone. The data, shown in Figure 6(b), were acquired in less than an hour in analogue conditions as for Au(111) shown in Fig. 5(d) (note the different energy and angular range for Au(111) and Bi$_2$Se$_3$ data).

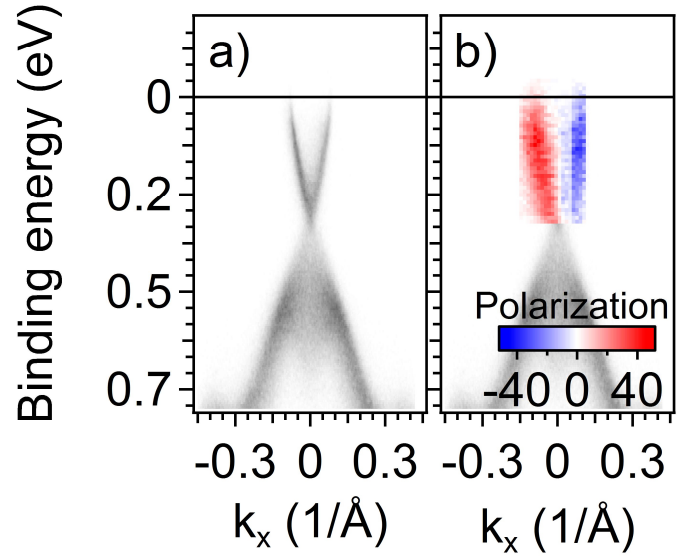

**Figure 6** (a) ARPES dispersion (binding energy versus $k_x$) of the topological surface state in the $Bi_2Se_3$ single crystal as measured along the K-Γ-K using hv=50 eV excitation photon energy; (b) the spin and angular-resolved data (colored scale showing the spin polarization) superimposed on the ARPES data of Fig. 6(a) reflect the spin texture of the topological surface state Dirac cone.

## 4. Data management

The VESPA data acquisition has been integrated within the SES program that drives the DA30 analyzer. The control of the target magnetization routine is also integrated in the data acquisition software via a separate dynamic link library (dll) developed by APE-NFFA. The data are saved either as generic text files, or in the Igor Wavemetrics format for facilitated data analysis. The procedures developed for the data analysis within Igor are available to the users for on-line as well as remote analysis.

## 5. Conclusion

A state of the art SPARPES spectrometer has been implemented under the NFFA-Trieste project and is fully operational since the end of 2015 as a users facility at the beamline APE on the Elettra storage ring, accessible also via the NFFA-Europe ([www.nffa.eu](www.nffa.eu)) and NFFA-Trieste ([www.trieste.nffa.eu](www.trieste.nffa.eu)) facilities. The efficient (FOM=$3.3 \cdot 10^{-2}$, S = 0.5) twin-VLEED spin detector VESPA allows for 3D spin vector measurements of high energy and momentum resolved ARPES data as obtained with polarization controlled synchrotron radiation in the far UV energy range. *Complete Photoemission Experiments* are made possible as data of adequate statistics can be collected in a short time allowing for spectroscopy and sample treatments during a typical beamtime access. The characteristics of VESPA along with the APE-LE beamline properties (ref. Panaccione *et al.*, 2009) and the in-situ facilities, that include the novel UHV Pulsed Laser Deposition growth chamber (Orgiani *et al.*, 2017), make the new APE-NFFA laboratory a very advanced facility for the study of highly correlated materials, surfaces and nanostructured systems.

**Acknowledgements** Chiara Bigi acknowledges support by NOXSS PRIN (2012Z3N9R9) Project of MIUR, Italy; this work has been performed in the framework of the Nanoscience Foundry and Fine Analysis (NFFA-MIUR Italy) facility project. It also represents part of the Tesi Magistrale of Chiara Bigi (Milano, 2016). We thank T. Pincelli for helpful and valuable discussions.